 \documentclass[a4paper,12pt]{article}\newcommand{\affiliation}[1]{}

\def\affiliation#1{\gdef\@affiliation{#1}}
\gdef\@affiliation{}
\newcommand{\Ord}[1]{{\cal O}\left(#1\right)}
\newcommand{\btau}{\mbox{\boldmath$\tau$}}
\newcommand{\frc}[2]{\mbox{$\frac{ #1}{ #2}$}}

\begin{document}
\title{ The Accurate Modelling of Thin 3D Fluid Flows with Inertia on
            Curved Substrates }
\author{Zhenquan Li and A.J. Roberts
\\Dept of Mathematics \& Computing\\
  University of Southern Queensland\\
 Toowoomba, Queensland 4350, Australia}
\maketitle
\markboth{Z. Li \& A.J. Roberts}{The Accurate Modelling of 3D Fluid Flow}

\section{Introduction}

Mathematical models and numerical simulations for thin-film flows of a
fluid have important applications in industrial and natural processes
\cite{Ruschak85}, \cite{Roskes69}, \cite{Schwartz95a},
\cite{Schwartz95b}, \cite{Chang94}, \cite{Grotberg94},
\cite{Moriarty91}.  Herein, we consider the slow motion of a thin
liquid layer of an incompressible, Newtonian fluid over an arbitrary
solid, stationary curved substrate.  In a three dimensional and very
slow lubrication flow, a model for the evolution of a film on a curved
substrate is shown \cite{Roy96} to be
\begin{equation}
\frac{\partial \zeta}{\partial t}\approx-\frac{1}{3}{\nabla}
     \cdot\left[\eta^2\zeta{\nabla }\tilde{\kappa}
      -\frac{1}{2}\eta^4(\kappa {\bf I}-{\bf K})
     \cdot{\nabla }\kappa\right]\,,
     \label{Elubr}
\end{equation}
where $\zeta=\eta-\frac{1}{2}\kappa\eta^2+\frac{1}{3}k_1k_2\eta^3$ is
proportional to the amount of fluid locally ``above'' the substrate;
$\tilde{\kappa}$ is the mean curvature of the free surface of the
film; {\bf K} is the curvature tensor of the substrate; $k_1$, $k_2$,
and $\kappa=k_1+k_2$ are the principal curvatures and the mean
curvature of the substrate respectively; and the operator $\nabla$ is
defined in a coordinate system of the curved substrate as introduced
in Section~2. 
This model accounts for the curvature of the substrate and that of the
surface of the film.
However, in many applications this model of slow flow of a thin fluid
film has limited usefulness; instead a model expressed in terms of
both the fluid layer thickness and the lateral velocity is needed to
resolve faster wave-like dynamics \cite[p110]{Chang94}.
Roberts \cite{Roberts97} derives such a model for two dimensional flow.
Here, based upon the Navier-Stokes equations for a viscous fluid,
Section~3, we derive the following model for three
dimensional flow:
\begin{eqnarray}
\frac{\partial \eta}{\partial t}&\!\!=\!\!&
-{\nabla}\cdot(\eta\bar{\bf u})\,,\label{Einert}
\\\nonumber
{\cal R}\frac{\partial \bar{\bf u}}{\partial t} &\!\!\approx\!\!&
-\frac{\pi^2}{4}\frac{\bar{\bf u}}{\eta^2}
+\frac{\pi^2}{12}\left({\nabla}\kappa+{\bf g}_s\right)
-\left(2{\bf K}+\kappa{\bf I}\right)\frac{\bar{\bf u}}{\eta}\,.
\end{eqnarray}
where ${\cal R}$ is a Reynolds number of the flow; ${\bf g}_s$ is the
component of gravity tangential to the substrate; and $\bar{\bf
u}=\left(\bar{u}_1,\bar{u}_2\right)$ is the depth-averaged average
lateral velocity.
The first equation is a direct consequence of the conservation of
fluid.
The model's second equation incorporates inertia, viscous drag on the
substrate, surface tension forcing caused by gradients of curvature,
gravitational forcing, and the geometric complexity of the substrate,
respectively.

The asymptotic accuracy of this model is assured by a systematic
derivation based upon centre manifold theory~\cite{Carr81} as applied
in Section~4.
The physical fields associated with the above model are deduced as
part of the procedure; approximations of the velocity and pressure
fields are recorded in Section~5. 
The model~(\ref{Einert}) is very general: it encompasses all substrate
shapes and contains the lubrication model simply by setting $\cal R$
to zero and substituting for $\bar{\bf u}$ in the first equation.
This model describes the dynamics of a very wide range of thin fluid
flows.

\section{The Orthogonal Curvilinear Coordinate System}

Let $\cal S$ denote the solid substrate.  If $\cal S$ has no umbilical
point, i.e., there is no point on $\cal S$ at which two principal
curvatures coincide, then there are exactly two mutually orthogonal
principal directions in the tangent plane at every point in ${\cal S}$
\cite[Theorem 10-3]{Gugg63}.  Let ${\bf e}_1$ and ${\bf e}_2$ be the
unit vectors in these principal directions, and ${\bf e}_3$ the unit
normal to the substrate in the side of fluid flow.  These basis unit
vectors determine a curvilinear orthonormal coordinate system
$(x_1,x_2,y)$.  Such a coordinate system is called a \emph{Darboux}
frame \cite{Gugg63}. The corresponding metric coefficients of the
coordinate system are
\begin{eqnarray}
h_i=m_i(1-k_iy)\,,\quad h_3=1\,.
\nonumber
\end{eqnarray}

Let $y=\eta(t,x_1,x_2)$ describe the free surface of the fluid.  As
derived by Roy {\it et al} \cite[eqn(37)]{Roy96}, the mean curvature
of the free-surface is
\begin{eqnarray}
&\!\!\!\!\!&\tilde{\kappa}=\frac{1}{\tilde{h}_1\tilde{h}_2}\left[\frac{\partial}{\partial x_1}
\left(\frac{\tilde{h}_2^2\eta_{x_1}}{\cal A}\right)+\frac{\partial}
{\partial x_2}
\left(\frac{\tilde{h}_1^2\eta_{x_2}}{\cal A}\right)\right]\quad\quad
\nonumber\\&\!\!\!\!\!&
+\frac{1}{\cal A}\left[\left(\tilde{h}_1^2
+{\eta_{x_1}}^2\right)\frac{m_2k_2}{\tilde{h}_1}
+\left(\tilde{h}_2^2+{\eta_{x_2}}^2\right)\frac{m_1k_1}{\tilde{h}_2}\right]
\nonumber
\end{eqnarray}
where $\tilde{h}_i=m_i(1-k_i \eta)$ are the metric coefficients
evaluated on the free surface and
\begin{center}
${\cal A}=\sqrt{\tilde{h}_1^2\tilde{h}_2^2+\tilde{h}_2^2\eta_{x_1}^2
+\tilde{h}_1^2\eta_{x_2}^2}$
\end{center}
is proportional to the free-surface area above a patch $dx_1\times
dx_2$ of the substrate.

We consider that the spatial derivatives of the fluid flow and the
curvatures of the substrate are much smaller than the metric
coefficients of the coordinate system, so that an approximation to
$\tilde{\kappa}$, as needed for (\ref{Elubr}), is
\begin{displaymath}
\tilde{\kappa}=\nabla^2\eta+\frac{k_1}{1-k_1\eta}+\frac{k_2}{1-k_2\eta}
+\Ord{\kappa^3+\nabla^3\eta}\,,
\end{displaymath}
where in the substrate coordinates
\begin{displaymath}
\nabla^2\eta=\frac{1}{{m}_1 {m}_2}\left[\frac{\partial}{\partial x_1}
              \left(\frac{{m}_2}{{m}_1}\frac{\partial \eta}{\partial
x_1}\right)+
              \frac{\partial } {\partial x_2}\left(\frac{{m}_1}{{m}_2}
              \frac{\partial \eta}{\partial x_2}\right)\right].
\end{displaymath}
For later use, also observe that on the free-surface, unit tangent
vectors $\tilde{\bf t}_1$, $\tilde{\bf t}_2$ and the unit normal
vector $\tilde{\bf n}$ are
\begin{eqnarray*}
\tilde{\bf t}_i&=&(\tilde{h}_i{\bf e}_1+\eta_{x_i}{\bf e}_3)
/\sqrt{{\tilde{h}_i}^2+\eta_{x_i}^2}\,,
\\
\tilde{\bf n}&=&\frac{-\tilde{h}_2\eta_{x_1}{\bf e}_1
-\tilde{h}_1\eta_{x_2}{\bf e}_2
+\tilde{h}_1\tilde{h}_2{\bf e}_3}
{\sqrt{(\tilde{h}_2\eta_{x_1})^2
+(\tilde{h}_1\eta_{x_2})^2
+(\tilde{h}_1\tilde{h}_2)^2}}\,.
\end{eqnarray*}

\section{Equations of Motion and Boundary Conditions}

Consider the Navier-Stokes equations for an incompressible fluid flow
moving with velocity field ${\bf u}=(u_1,u_2,v)$ and pressure field
$p$.
The flow dynamics are driven by pressure gradients along the substrate
and caused by both surface tension forces, coefficient $\sigma$,
varying due to variations of the curvature of the free surface of the
fluid, and a gravitational body force, {\bf g}, of magnitude $g$ in
the direction of the unit vector $\hat{\bf g}$.
Let the reference length be a characteristic thickness of the film
$H$, the reference time $\mu H/\sigma$, the reference velocity
$U=\sigma/\mu$, and the reference pressure $\sigma/H$.  Then the
non-dimensional incompressible and Navier-Stokes equations are:
\begin{displaymath}
\nabla\cdot{\bf u} = 0\,,
\end{displaymath}
\begin{displaymath}
{\cal R}\left[\frac{\partial {\bf u}}{\partial t}
+{\bf u}\cdot\nabla{\bf u}\right]
 =  -\nabla p+\nabla^2{\bf u}+b\hat{\bf g}\,,
\end{displaymath}
where ${\cal R}=\sigma\rho H/\mu^2$ is a Reynolds number
characterising the importance of the inertial terms---it may be
written as $UH/\nu$ for above reference velocity---and $b=\rho g
H^2/\sigma$ is a Bond number characterising the importance of the
gravitational body force.

We asymptotically solve these non-dimensional continuity and
Navier-Stokes equations in the curvilinear coordinate system described
in Section~2 with the following boundary conditions
\begin{enumerate}
\item
The fluid does not slip along the stationary substrate, that is
\begin{displaymath}
{\bf u= 0}\quad\mbox{on $y=0$}\,.
\end{displaymath}
\item
The fluid satisfies the kinematic free surface boundary condition
\begin{displaymath}
\frac{\partial \eta}{\partial t}
=v-\frac{u_1}{\tilde{h}_1}\frac{\partial \eta}{\partial x_1}
-\frac{u_2}{\tilde{h}_2}\frac{\partial \eta}{\partial x_2}
\quad\mbox{on $y=\eta$}\,.
\end{displaymath}
\item
The stress across the free surface is caused by surface tension, in
non-dimensional form
\begin{displaymath}
-p\tilde{\bf n}+\tilde{\btau}\cdot\tilde{\bf n}
=\tilde{\kappa}\tilde{\bf n}\quad\mbox{on $y=\eta$}\,,
\end{displaymath}
where $p$ is the fluid pressure relative to the assumed zero pressure
of the medium above, and $\tilde{\btau}$ is the fluid's deviatoric
stress tensor evaluated at the free surface.
\end{enumerate}

\section{A Centre Manifold Basis for the Model}

Using centre manifold techniques \cite{Carr81}, we derive a
low-dimensional model of the dynamics described in detail by the
non-dimensional continuity and Navier-Stokes equations in the general
curvilinear coordinate system.
We assume that the spatial extent of the fluid flow and the curving of
the substrate occur on a much larger scale than the thickness of the
fluid.
Thus, $\epsilon$ is introduced to parameterise the relatively small
effect of these spatial variations and curvature, i.e., we introduce
re-scaled * variables
\begin{eqnarray}
\frac{\partial}{\partial x_i}
=\epsilon\frac{\partial}{\partial x^{\ast}_i}\,,\quad
k_i=\epsilon k_i^*\,,\quad
\kappa=\epsilon{\kappa}^*\,.
\nonumber
\end{eqnarray}
To justify treating the gravitational forcing as small, we may
introduce a parameter $\beta$ such that the Bond number $b=\beta^2$;
then after adjoining $\frac{\partial \beta}{\partial t}=0$ to the
dynamical equations centre manifold theory justifies an asymptotic
expansion in $\beta$.
As introduced by Roberts \cite{Roberts96b,Roberts97}, we also modify
the tangential stress on free surface in order to establish three
critical modes in the centre manifold model rather than the one
natural critical mode of lubrication theory.
The modification is parameterised by $\gamma$; we construct a centre
manifold for asymptotically small $\gamma$; then evaluating the
results for $\gamma=1$ recovers a model for the physical dynamics.
Previous work \cite{Roberts96b} has shown that such evaluation of
these low-order expansions at $\gamma=1$ is accurate.

For convenience we drop hereafter the ``$\ast$'' superscript on all
re-scaled variables.  The non-dimensional continuity and Navier-Stokes
equations in the curvilinear coordinate system become the following
governing equations (where $i'=3-i$):
\begin{displaymath}
\epsilon\frac{\partial}{\partial x_1}(h_2 u_1)
+\epsilon\frac{\partial }{\partial x_2}(h_1 u_2)+
\frac{\partial }{\partial y}(h_1h_2 v)=0\,,
\end{displaymath}
\begin{eqnarray*}  \\&&
{\cal R}\left[\frac{\partial u_i}{\partial t}
+\epsilon\frac{u_1}{h_1}\frac{\partial u_i} {\partial x_1}
+\epsilon\frac{u_2}{h_2}\frac{\partial u_i}{\partial x_2}
+v\frac{\partial u_i} {\partial y}\right.
\\&&
+\left.\epsilon\frac{u_{i'}}{h_ih_{i'}}
\left(u_i\frac{\partial h_i}{\partial x_{i'}}
-u_{i'}\frac{\partial h_{i'}}{\partial x_i}\right)
-\epsilon m_ik_i\frac{u_iv}{h_i}\right]
\\&&=
-\frac{\epsilon}{h_i}\frac{\partial p}{\partial x_i}
+\frac{1}{h_1 h_2}\left[\epsilon^2\frac{\partial }{\partial x_1}
\left(\frac{h_2}{h_1}\frac{\partial u_i}{\partial x_1}\right)\right.
\\&&
+\left.\epsilon^2\frac{\partial }{\partial x_2}
\left(\frac{h_1}{h_2}\frac{\partial u_i}{\partial x_2}\right)
+\frac{\partial }{\partial y}
\left({h_1}{h_2}\frac{\partial u_i}{\partial y}\right)\right]
+b\hat g_i\,,
\end{eqnarray*}
\begin{eqnarray*} \\&\!\!\!&
{\cal R}\left[\frac{\partial v}{\partial t}
+\epsilon\frac{u_1}{h_1}\frac{\partial v} {\partial
x_1}+\epsilon\frac{u_2}{h_2}\frac{\partial v}{\partial x_2}
+v\frac{\partial v} {\partial y}+\epsilon m_1k_1\frac{u_1^2}{h_1}\right.
\\&\!\!\!&
+\left.\epsilon m_2k_2\frac{u_2^2}{h_2}\right]
=-\frac{\partial p}{\partial y}
+\frac{1}{h_1 h_2}\left[\epsilon^2\frac{\partial }{\partial x_1}
\left(\frac{h_2}{h_1}\frac{\partial v}{\partial x_1}\right)\right.
\\&\!\!\!&
+\left.\epsilon^2\frac{\partial }{\partial x_2}
\left(\frac{h_1}{h_2}\frac{\partial v}{\partial x_2}\right)
+\frac{\partial }{\partial y}
\left({h_1}{h_2}\frac{\partial v}{\partial y}\right)\right]
+b\hat g_3\,,
\end{eqnarray*}
where the re-scaled scale factors are $h_i=m_i(1-\epsilon k_i
y)$, and the boundary conditions become
\begin{displaymath}
{\bf u= 0}\quad\mbox{on $y=0$}\,,
\end{displaymath}
\begin{displaymath}
\frac{\partial \eta}{\partial t}
=v-\epsilon\frac{u_1}{\tilde{h}_1}\frac{\partial \eta}{\partial x_1}
-\epsilon\frac{u_2}{\tilde{h}_2}\frac{\partial \eta}{\partial x_2}
\quad\mbox{on $y=\eta$}\,,
\end{displaymath}
\begin{displaymath}
\tilde{\bf t}_i\cdot\tilde{\btau}\cdot\tilde{\bf n}=
(1-\gamma)\frac{m_im_1m_2u_i}{\eta l_i l}
\quad\mbox{on}\quad y=\eta\,,
\end{displaymath}
\begin{displaymath}
\tilde{\bf n}\cdot\tilde{\btau}\cdot\tilde{\bf n}=p+\tilde{\kappa}
\quad\mbox{on}\quad y=\eta\,,
\end{displaymath}
where
\begin{eqnarray}
l_i&=&\sqrt{{\tilde{h}_i}^2+{\epsilon^2\eta_{x_i}^2}}\,,\nonumber\\
l&=&\sqrt{(\epsilon\tilde{h}_2\eta_{x_1})^2+(\epsilon\tilde{h}_1\eta_{x_2})^2
      +(\tilde{h}_1\tilde{h}_2)^2}\,,\nonumber
\end{eqnarray}
and unit tangent vectors $\tilde{\bf t}_1$, $\tilde{\bf t}_2$ and the
unit normal vector $\tilde{\bf n}$ are
\begin{eqnarray}
\tilde{\bf t}_i&=&(\tilde{h}_i{\bf e}_i+\epsilon\eta_{x_i}{\bf e}_3)/{l_i}\,,
\nonumber\\
\tilde{\bf n}&=&(-\epsilon\tilde{h}_2\eta_{x_1}{\bf e}_1
-\epsilon\tilde{h}_1\eta_{x_2}{\bf e}_2
+\tilde{h}_1\tilde{h}_2{\bf e}_3)/l\,.
\nonumber
\end{eqnarray}

Then by adjoining the trivial dynamical equations
\begin{displaymath}
\frac{\partial \epsilon}{\partial t}=0\,,
\quad \frac{\partial \gamma}{\partial t}=0
\quad\mbox{and}\quad \frac{\partial \beta}{\partial t}=0\,,
\end{displaymath}
we get a new dynamical system in the variables {\bf u}, $\eta$, $p$,
$\epsilon$, $\gamma$ and $\beta$.  The original system will be
recovered by setting $\epsilon=1$, $\gamma=1$, $\beta=\sqrt{b}$.
However, the two systems are quite different from the view of centre
manifold theory.  Theory \cite{Carr81} justifies treating all terms
that are multiplied by the three introduced parameters as nonlinear
perturbating effects in the new system.

Then the linear part (in this new sense) of the governing equations
and boundary conditions have solutions: $u_i=v=p=0$\,,
$\eta=\mbox{constant}$; and independently $v=p=0$, $u_i\propto
\sin(\omega y/\eta)\exp(\lambda t)$ where
\begin{displaymath}
\lambda=-\frac{\omega^2}{{\cal R}\eta^2}\,,
\quad\mbox{such that}\quad \omega=\tan\omega\,,
\end{displaymath}
Thus, the critical modes (with zero eigenvalue) are associated with
varying thickness $\eta$, and independently with the two shear modes
$u_i \propto y$.  The system we consider has three critical modes and
three trivial parameter modes, all other modes decay exponentially
quickly. Also, the nonlinear terms are continuous at least.  Therefore a
low-dimensional model of the system is justifiably constructed by
centre manifold theory.

Denote the variables in the original system by ${\bf
v}(t)=(\eta,u_1,u_2,v,p)$.
Centre manifold theory guarantees that there exist functions {\bf V}
and {\bf G} respectively describing the shape of the centre manifold
and the evolution thereon, namely
\begin{displaymath}
{\bf v}(t)={\bf V}(\eta,\bar{u}_1,\bar{u}_2)\,,
\end{displaymath}
\begin{displaymath}
\mbox{such that}\quad
\frac{\partial}{\partial t}\left[\begin{array}{c} \eta \\
\bar{u}_1 \\ \bar{u}_2\end{array} \right]
={\bf G}(\eta,\bar{u}_1,\bar{u}_2)\,,
\end{displaymath}
where dependence upon the constant parameters
$(\epsilon,\gamma,\beta)$ is implicit in the above, and $\bar{u}_i$
are depth-averaged velocities measuring the amplitude of the shear
modes $u_i\propto y$.
The aim now is to find functions {\bf V} and {\bf G} such that ${\bf
v}(t)$ are actual solutions of governing equations.
We calculate {\bf V} and {\bf G} by an iteration using computer
algebra \cite{Roberts96a}.
Suppose that an approximation $\tilde{\bf V}$ and $\tilde{\bf G}$ has
been calculated and that we seek corrections ${\bf V'}$ and ${\bf G'}$.
Substituting
\begin{displaymath}
{\bf v}=\tilde{\bf V}+{\bf V'}\,,
\quad\mbox{such that}\quad
\frac{\partial}{\partial t}
\left[\begin{array}{c} \eta \\ \bar{u}_1 \\ \bar{u}_2\end{array} \right]
=\tilde{\bf G}+{\bf G'}
\end{displaymath}
into the governing equations then rearranging, dropping products of
corrections, and using the linear approximation wherever terms
multiply corrections, we obtain a system of equations for the
corrections which is in the homological form
\begin{displaymath}
{\cal L}{\bf V'}+A{\bf G'}=\tilde{\bf R}\,,
\end{displaymath}
where ${\cal L}$ is the linear part of the governing equations and the
boundary conditions, $A$ is a matrix, and $\tilde{\bf R}$ is the
residual of the governing equations using the reigning approximations,
$\tilde{\bf V}$ and $\tilde{\bf G}$.
The procedure for solving the equations is as follows: first, choose
${\bf G'}$ such that $\tilde{\bf R}-A{\bf G'}$ is in the domain of
${\cal L}$; second, solve ${\cal L}{\bf V'}=\mbox{rhs}$ with the given
boundary conditions; then regard $\tilde{\bf V}+{\bf V'}$ and
$\tilde{\bf G}+{\bf G'}$ as the new approximations of $\tilde{\bf V}$
and $\tilde{\bf G}$ respectively.
Repeat the procedure until the residual $\tilde{\bf R}$ becomes zero
to the required order of error.
Then the low-dimensional model has the same order of error by the
Approximation Theorem \cite{Carr81} in centre manifold theory.
A computer algebra program (obtainable from the authors) is run to
perform the computations.
The key to the correctness of the results we report is the correct
coding of the residuals within the body of the iteration.

\section{The Low-dimensional Model}

First, computing to low-order in the small parameters gives the
following fields in terms of the parameters
\begin{eqnarray}
p&=&-\kappa\epsilon-{\nabla}^2\eta\epsilon^2
   -\eta\kappa_2\epsilon^2
   +\eta(Y-1)g_3\epsilon^2
\nonumber\\
   & &-\eta^{-1}\bar{\bf u}\cdot\nabla\eta
   \left(\frc{1}{2}\gamma-\frc{1}{2}\gamma Y +2-2 Y\right)\epsilon^2
\nonumber\\
   & &-\nabla\cdot\bar{\bf u}
     (2+2Y-\frc{3}{2}\gamma+\frc{1}{2}\gamma Y)\epsilon^2\nonumber\\
& &+\Ord{\epsilon^3+\bar{u}^3+\beta^3,\gamma^2},
\nonumber
\\
{\bf u}_s&=&\bar{\bf u}\left(2Y+\frc{1}{2}\gamma Y-\gamma Y^3\right)\epsilon
   \nonumber\\
   & &+\eta^2\left(\nabla\kappa+{\bf g}_s\right)\left(\frc{5}{24} Y
     -\frc{1}{2} Y^2+\frc{1}{4} Y^3\right.\nonumber\\
   & &-\left.\frc{17}{480}\gamma Y+\frc{23}{240}\gamma Y^3
   -\frc{3}{80}\gamma Y^5\right)\epsilon^2
   \nonumber\\
   & &-\eta\kappa\bar{\bf u}\left(\frc{5}{12} Y-Y^2
    +\frc{1}{2} Y^3+\frc{19}{120}\gamma Y
    \right.\nonumber\\
   & &\left.-\frc{1}{4}\gamma Y^2
   -\frc{17}{60}\gamma Y^3+\frc{1}{2}\gamma Y^4
   -\frc{3}{20}\gamma Y^5\right)\epsilon^2
   \nonumber\\&&
    +\eta{\bf K}\cdot\bar{\bf u}\left(\frc{1}{2} Y- Y^3
    -\frc{1}{4}\gamma Y
   -\frc{1}{10}\gamma Y^3
   \right.\nonumber\\&&\left.
   +\frc{3}{10}\gamma Y^5\right)\epsilon^2
   +\Ord{\epsilon^3+\bar{u}^3+\beta^3,\gamma^2},
\nonumber\\
v&=&\bar{\bf u}\cdot\nabla\eta
\left(Y^2+\frc{1}{4}\gamma Y^2-\frc{3}{4}\gamma Y^4\right)\epsilon^2
 \nonumber\\
  & &-\eta\nabla\cdot\bar{\bf u}
  \left(Y^2+\frc{1}{4}\gamma Y^2-\frc{1}{4}\gamma Y^4\right)\epsilon^2
\nonumber\\
  & &+\Ord{\epsilon^3+\bar{u}^3+\beta^3,\gamma^2},
\nonumber
\end{eqnarray}
where $Y=y/\eta$, $\bar{u}=|\bar{\bf u}|$, ${\bf u}_s=(u_1,u_2)$, and
$\Ord{\epsilon^p+\bar{u}^q+\beta^m,\gamma^n}$ is used to denote terms
$s$ which satisfy that $s/(\epsilon^p+\bar{u}^q+\beta^m)$ is bounded
as $(\epsilon,\bar{u},\beta)\rightarrow$0, or $s/\gamma^n$ is bounded
as $\gamma\rightarrow$0.
In the expression for $p$, the first and second terms are effects of
the substrate curvature and free-surface, the third term is the
correction for the first, the fourth term is hydrostatic, and the
others are effects due to the motion of fluid.
In the expression for $\bar{\bf u}$, the first line is the basic shear
profile modified by boundary conditions, the second and third lines
are modification due to forcing, and the others are effects of
curvature.
The expression for $v$ expresses the vertical component of velocity is
only dependent of the variations of free-surface and the other
components.

The corresponding evolution on this centre manifold is then
\begin{eqnarray}
\frac{\partial \eta}{\partial t}&=&-\epsilon^2\
{\nabla}\cdot(\eta\bar{\bf u})
+\Ord{\epsilon^3+\bar{u}^3+\beta^3,\gamma^2},
\nonumber
\\
{\cal R}\frac{\partial \bar{\bf u}}{\partial t}&=&
\left({\nabla}\kappa+{\bf
g}_s\right)\left(\frc{3}{4}+\frc{1}{10}\gamma\right)\epsilon
         \nonumber\\
& &+\eta^{-1}\kappa\bar{\bf u}\left(\frc{3}{5}\gamma-\frc{3}{2}\right)\epsilon
\nonumber\\
 & &+\eta^{-1}{\bf K}\cdot\bar{\bf u}\left(\frc{6}{5}\gamma-3\right)\epsilon
\nonumber\\
    & &-3\eta^{-2}\bar{\bf u}\gamma
    \nonumber\\
& &+\Ord{\epsilon^3+\bar{u}^3+\beta^3,\gamma^2}\,.
\nonumber
\end{eqnarray}
The first equation gives the general expression representing mass
conservation for fluid considered. The first line in the right-hand
side of the second equation is forcing, the second and third lines are
curvature effects and the fourth line is drag.

To recover a model of the original dynamics, we need to set $\gamma=1$.
But as is apparent from the above, every coefficient in the centre
manifold model is a power series in $\gamma$.
From earlier computation in \cite{Roberts96b}, we estimate that the
radii of convergence of such $\gamma$ series are much greater than 1.
Thus after computing to higher order in $\gamma$, we calculate every
coefficient in the model from the first five terms in $\gamma$ series
by setting $\gamma=1$.
We also need to set $\epsilon=1$ which is valid provided the
length-scales in the resolved dynamics are indeed much larger than the
fluid thickness, that is, if the gradients are small enough.
The low-dimensional model then becomes as given in the Introduction
by~(\ref{Einert}) with errors $\Ord{\nabla^3+\bar{u}^3+\beta^3}$.

\section{Conclusion}

 This model conserves fluid and accurately accounts for the effects of the
 curvature of substrate, surface tension, gravitational forcing and fluid
 inertia.

 The low-dimensional model given in this paper is a simpler version. One may
 adjust a dynamical model to suit a particular application.

\end{document}